\documentclass[12pt]{JHEP3}

\def\cee{{\relax\hbox{$\inbar\kern-.3em{\rm C}$}}}

\newcommand{\be}{\begin{equation}}
\newcommand{\ee}{\end{equation}}
\newcommand{\bea}{\begin{eqnarray}}
\newcommand{\eea}{\end{eqnarray}}
\newcommand{\bml}{\begin{mathletters}}
\newcommand{\eml}{\end{mathletters}}
\newtheorem{theorem}{Theorem}

\newtheorem{proposition}{Proposition}

\def\darr#1{\raise1.5ex\hbox{$\leftrightarrow$}\mkern-16.5mu #1}

\def\roughly#1{\raise.3ex\hbox{$#1$\kern-.75em\lower1ex\hbox{$\sim$}}}

\def\IB{\relax\hbox{$\inbar\kern-.3em{\rm B}$}}
\def\IC{\relax\hbox{$\inbar\kern-.3em{\rm C}$}}
\def\ID{\relax\hbox{$\inbar\kern-.3em{\rm D}$}}
\def\IE{\relax\hbox{$\inbar\kern-.3em{\rm E}$}}
\def\IF{\relax\hbox{$\inbar\kern-.3em{\rm F}$}}
\def\IG{\relax\hbox{$\inbar\kern-.3em{\rm G}$}}
\def\IGa{\relax\hbox{${\rm I}\kern-.18em\Gamma$}}
\def\IH{\relax{\rm I\kern-.18em H}}
\def\IK{\relax{\rm I\kern-.18em K}}
\def\IL{\relax{\rm I\kern-.18em L}}
\def\IP{\relax{\rm I\kern-.18em P}}
\def\IR{\relax{\rm I\kern-.18em R}}
\def\IZ{\relax\ifmmode\mathchoice{
\hbox{\cmss Z\kern-.4em Z}}{\hbox{\cmss Z\kern-.4em Z}}
{\lower.9pt\hbox{\cmsss Z\kern-.4em Z}} {\lower1.2pt\hbox{\cmsss
Z\kern-.4em Z}} \else{\cmss Z\kern-.4em Z}\fi}
\def\II{\relax{\rm I\kern-.18em I}}

\def\ee#1{{\rm erf}\left(#1\right)}

\def\inbar{\,\vrule height1.5ex width.4pt depth0pt}

\font\cmss=cmss10 \font\cmsss=cmss10 at 7pt

\def\lref{\begingroup\obeylines\lr@f}
\def\lr@f#1#2{\gdef#1{\ref#1{#2}}\endgroup\unskip}

\def\and{{a^\dagger_n}}

\def\math@note#1{\gdef\@eqnlabel{LAB: #1}}
\title{ORBIFOLD COMPACTIFICATION AND SOLUTIONS
OF M--THEORY FROM MILNE SPACES}
\author{A. A. BYTSENKO \\
Departamento de F\'{\i}sica, Universidade Estadual de
Londrina\\
Caixa Postal 6001, Londrina-Paran\'a, Brazil}
\author{M. E. X. GUIMAR\~AES \\
Departamento de Matem\'atica, Universidade de Bras\'{\i}lia \\
Campus Universit\'ario, Bras\'{\i}lia-DF, Brazil}
\author{R. KERNER \\
Laboratoire de Physique Th\'eorique des Liquides (UMR 7600)\\
Universit\'e Pierre et Marie Curie, Tour 22, 4-\'eme \'etage,
Bo\^ite 142 \\
4, Place Jussieu, 75005 Paris, France}

\abstract{In this paper, we consider solutions and spectral
functions of M-theory from Milne spaces with extra free dimensions.
Conformal deformations to the metric associated with the real
hyperbolic space forms are derived.
For the three-dimensional case, the orbifold identifications
$SL(2,{\mathbb Z}+i{\mathbb Z})/\{\pm Id\}$, where $Id$ is the
identity matrix, is analyzed in
detail. The spectrum of a eleven-dimensional field theory can be
obtained with the help of the
theory of harmonic functions in the fundamental domain of this
group and it is associated with the cusp forms and the Eisenstein
series. The supersymmetry surviving for supergravity solutions
involving real hyperbolic space factors is briefly discussed.}


\begin{document}

\section{Introduction}

The true nature of gravity, though well measured for solar system
scale experiments, is not well determined at larger or smaller
scales. For example, the assumptions of dark matter and/or energy
are made in order to fit the observed Universe within Einstein
theory, yet it is quite possible that it is the theory of gravity
which should be modified to accomodate these data. Theoretically,
the possibility that gravity might not be fundamentally described
by a purely tensorial theory in four dimensions is growing in
importance. This is in part a consequence of superstring theory,
which is consistent in ten dimensions (or M-theory in eleven
dimensions), but also the more phenomenological recent
developments of ``braneworld'' cosmological scenarios
\cite{ADD,ADD1,Antoniadis,RS,RS1} have motivated the study of other
gravitational theories in four-dimensions.

It is generally accepted that M-theory may provide a
consistent quantum theory of gravity. Nevertheless, it is
understood by now that to insert this theory in time-dependent
backgrounds can bring a number of technical problems such as the
appearance of closed timelike curves and the spacetime resulting
from string compactification is not Hausdorff. Yet, open questions
in cosmology such as the initial (big bang) singularity and the
initial boundary conditions remain a challenge and they are the
main motivations to consider string cosmology.

Initial boundary conditions and the requirement of homogeneity for
the cosmological solution imply that the geometry has a form of
higher dimensional Milne universe along a null hypersurface, with
negative constant curvature in the spatial sector. This spacetime
can be viewed particularly as hyperbolic compactifications in
M-theory (see \cite{Kehagias,Russo,Bytsenko00}), which have recently
attracted some
interest as they lead to interesting cosmologies \cite{Townsend}.
Cosmological string models in Milne universe have been considered
by many authors. Milne spaces in the context of inflationary
cosmology were studied in \cite{Yamamoto,Tanaka}. String models in
(1+1)-dimensional Milne space were discussed in
\cite{Kehagias,Russo,Horowitz1,Khoury,Seiberg,Cornalba,Nekrasov}.
Discussions on
higher dimensional Milne spaces can be found in
\cite{Kehagias,Horowitz1}, and more recently  in
\cite{Russo,Cornalba}.

In the present paper we will extend these
previous works in order to contemplate the problem of hyperbolic
compactifications in the context of cosmological scenarios.
In particular, we will be interested in the general class of
time-dependent locally flat spacetimes obtained from the
$(N+1)-$dimensional Milne universe. Emphasis is devoted to
the $N=3$ case, which analyze in details and orbifold
identifications using the modular group
$\Gamma = SL(2,{\Bbb Z}+ i {\Bbb Z})/\{\pm Id\}$, where $Id$ is
the identity matrix is considered. In Sec. 2 we analyse
the class of conformal
deformations of Riemannian metrics and in particular the
conformal relation between Milne and hyperbolic space forms.
In Sec. 3 we take into account constant slice in Milne spaces.
In Sec. 4 we consider hyperbolic geometry in the spatial section
of Milne space. For co-compact groups $\Gamma$ (i.e. for
compact real hyperbolic manifolds) the heat kernel coefficients
are given in the explicit forms. Orbifolding
of the group $\Gamma$ we derived in explicit forms
the Selberg trace formula and the
determinant of Laplace type operators.
The study of this theory involves harmonic analysis on locally
symmetric spaces of rank one from which we will extract some results
for the brane picture.
Finally in Sec. 5 we discuss questions of supersymmetry surviving
under the orbifolding of a discrete group.

\section{Conformal deformations and Milne space forms}

Let $M$ be a $D=(N+1)-$dimensional Riemannian space with
metric
$
ds^2=g_{00}({\bf x})(dx^0)^2+g_{ij}({\bf x})dx^idx^j\:,
{\bf x}=\{x^j\}\,, i,j=1,...,N.
$
For the conformal deformations of the $g_{\mu \nu}$ the following
relation holds:
\begin{equation}
\widetilde{g}_{\mu\nu}({\bf x})=e^{2\sigma({\bf x})}g_{\mu\nu}
({\bf x})\,,\,\,\,\,\,\,\, \sigma({\bf x})\in C^{\infty}(M)
\mbox{.}
\label{conformal}
\end{equation}
Recalling that the partition function of field theory is given by
(in the Euclidean sector differential operators are elliptic)
$
W =\int d[\varphi]\,
\exp(-(1/2)\int_M d^Dx \varphi {\mathfrak L}\varphi)
$,
where $\varphi$ is a scalar density of wight $-1/2$ and
the operator ${\mathfrak L}$ has the form
$
{\mathfrak L}=-\triangle^g+m^2+\xi R^g
$,
where $m$ (the mass) and $\xi$ are arbitrary parameters, while
$\triangle^g$ and $R^g$ are respectively the Laplace-Beltrami operator
and the scalar curvature of the space with respect to the original
metric $g$. One gets the following result:

\begin{proposition} (A. A. Bytsenko, G. Cognola and S. Zerbini
\cite{Bytsenko3})\,\,\,
Let $\widetilde\varphi = e^{\sigma}\varphi$ then
the conformal deformations (\ref{conformal}) leads to:
${\widetilde {\mathfrak L}}=e^{-\sigma}{\mathfrak L}e^{\sigma}$, and
\begin{eqnarray}
R^{\widetilde g} & = & e^{-2\sigma}\left[R^g-2(D-1)\triangle^g
\sigma-(D-1)(D-2)g^{\mu\nu}
\partial_{\mu}\sigma\partial_{\nu}\sigma\right]\,
\nonumber \\
\triangle^{\widetilde g}\widetilde\varphi & = &
\frac{1}{4}e^{-\sigma}
\left[4\triangle^g
-2(D-2)\triangle^g\sigma
-(D-2)^2g^{\mu\nu}
\partial_{\mu}\sigma\partial_{\nu}\sigma
\right]\varphi
\nonumber \\
&=& e^{-\sigma}\left[\triangle^g+\xi_D(e^{2\sigma}R^{\widetilde g}-R^g)
\right]\varphi\,,
\nonumber \\
{\mathfrak L} & = & e^{\sigma}\{-\triangle^{\widetilde g}+
\xi R^{\widetilde g}
+e^{-2\sigma}[m^2+(\xi-\xi_D)R^g]\} e^{\sigma}
\mbox{,}
\label{}
\end{eqnarray}
where $\xi_D=(D-2)/4(D-1)$ is the conformal invariant factor.
\end{proposition}
The classical conformal invariance requires that the action $S$
is invariant in form, that is
$
\widetilde S = S[\widetilde\varphi,\widetilde g]
$,
(as to say $\widetilde {\mathfrak L} = {\mathfrak L}$).
As it is well known,
this happens only for conformally coupled massless fields ($\xi=\xi_D$).
For the partition function we have
$
\widetilde W = J[g,\widetilde g]\,W
$,
where $J[g,\widetilde g]$ is the Jacobian of the conformal
deformation.

For $0<t<1$ the asymptotic expansion holds
\begin{equation}
{\rm Tr} e^{-t{\mathfrak L}}
\simeq\sum_j A_j({\mathfrak L})t^{(j-D)/2}\,,
\,\,\,\,\,\,\,\,
A_j({\mathfrak L}) = (4\pi)^{-D/2}\int_{M} d^Dx
\sqrt{g} a_j(x|{\mathfrak L})
\mbox{,}
\label{kernel}
\end{equation}
where $a_j(x|{\mathfrak L})$ is the $j-$th Seeley-De Witt coefficient
(in the conformal invariant theories it is proportional to the
trace anomaly). If the boundary of a manifold is empty then
$A_j({\mathfrak L})=0$ for any odd $j$. The following results hold:

\begin{proposition} \,\,\,
Let us consider a family of conformal deformations
$
g^q_{\mu\nu}=e^{2q\sigma}g_{\mu\nu}
=e^{2(q-1)\sigma}\widetilde g_{\mu\nu}\,,
\sqrt{g^q}\equiv\sqrt{|{\rm det}\, g^q_{\mu\nu}|}=
e^{Dq\sigma}\sqrt{g}\,.
$
The metric is
$g_{\mu\nu}$ or $\widetilde{g}_{\mu\nu}$
according to whether $q=0$ or $q=1$ respectively.
Then,
\begin{eqnarray}
{\rm log}\, J[g_q,g_{q+\delta q}]
& = & {\rm log}\,\left[W_{q+\delta q}/W_q\right]
= (4\pi)^{-D/2}\delta q
\int_M d^Dx\sqrt{g^q} a_D(x|{\mathfrak L}^q)\sigma(x)\,,
\\
{\rm log}\, J[g,\widetilde{g}] & = & (4\pi)^{-D/2}
\int_0^1dq\int_M d^Dx\sqrt{g^q}\, a_D(x|{\mathfrak L}^q)\sigma(x)\,,
\label{lnJ}
\\
{\rm log}\,W & = & {\rm log}\,{\widetilde W}-{\rm log}\,
J[g,\widetilde g]
=\frac{d}{2ds}\zeta(s|\widetilde {\mathfrak L}\ell^2)|_{s=0}
-{\rm log}\,J[g,\widetilde{g}]\,.
\label{PF}
\end{eqnarray}
Eq. (\ref{PF}) has been derived with the help of the zeta-function
regularization, $\ell$ being an arbitrary parameter necessary to adjust
the dimensions.
\end{proposition}

\subsection{Remark}

The $(N+1)-$dimensional Milne space is described by the metric
$
ds^2=-dt^2+t^2 d{\Bbb H}^N\,,
$
where $d{\Bbb H}^N$ is the arc
element of the hyperboloid or upper half $N-$plane.
The space is flat, as it is
evident upon introducing Cartesian coordinates as follows:
$
U= ty^{-1},\, V=ty+U\sum_{j=1}^{N-1}
x_j^2,\,  X_j= Ux_j.
$
This provides the embedding
of the hyperboloid in $(N+1)-$Minkowski space,
$
ds^2=-dUdV+dX_j^2 \ ,
$
where the hyperboloid is described by
$
t^2 =
UV-\sum_{j=1}^{N-1}X_j^2\ ,
$
which exhibits the $SO_1(N,1)$ isometry of ${\Bbb H}^N$.

Before concluding this section, some remarks on the Milne metric
are in order. New coordinates in the Euclidean sector, $t\rightarrow it$,
can be introduced as follows: $\tau = {\rm log}\,t, t \neq 0$
($t=0$ is a harmless coordinate singularity and corresponds to a
horizon in this metric). This gives the new form for the metric:
$ds^2 = e^{2\tau}d\tau^2 + e^{2\tau}d{\mathbb H}^N$.
Taking into account
Eq. (\ref{conformal}) one can choose $\sigma ({\bf x}) = -\tau$.
In the Euclidean sector it gives
\begin{equation}
d{\widetilde s}^2=d\tau^2+d{\Bbb H}^N
\mbox{.}
\label{newmetric}
\end{equation}
Therefore, in a class of conformal deformations the metric
(\ref{newmetric}) is related to the initial metric of Milne space
and can be associated with spacetime forms of topology
$S^1\times{\mathbb H}^N$.
One can use angular coordinates and define the initial metric
as follows:
\begin{equation}
d{s}^2 = dt^2 + \frac{t^2}{\rho^2}
(d\rho^2+d\Omega^2_{N-1})
\:,
\label{rlo}
\end{equation}
where $d\Omega^2_{N-1}$ is the metric of a $(N-1)-$dimensional space.
The technique of the conformal deformations of the Rindler space
(except of a horizon) with its connection to a space with hyperbolic
spatial section has been discussed in \cite{Bytsenko3}. The metrics
of both spaces have the form
\begin{equation}
ds^2_{(Rindler)}
= \rho^2 dt^2 + d\rho^2 + d\Omega_{N-1}^2
\stackrel {\sigma = - {\rm log}\,\rho}{\Longrightarrow}
d{\widetilde s}^2_{(S^1\times{\mathbb H}^N)}
= dt^2 + \frac{1}{\rho^2}(d\rho^2 + d\Omega^2_{N-1})
\mbox{.}
\end{equation}
For the Milne space a similar deformation (except of a horizon)
in coordinate $\tau$ becomes
\begin{equation}
ds^2_{(Milne)} =  e^{2\tau}\left(d\tau^2 + \frac{1}{\rho^2}
(d\rho^2 + d\Omega_{N-1}^2)\right)
\stackrel {\sigma = - \tau}{\Longrightarrow}
d{\widetilde s}^2_{(S^1\times{\mathbb H}^N)} =
d\tau^2 + \frac{1}{\rho^2}(d\rho^2 + d\Omega^2_{N-1})
\mbox{.}
\end{equation}
The metrics on Rindler and Milne spaces are in the conformal class,
and connections between their conformal deformations reads
$\tau = {\rm log}\,\rho$. Here we derive the operator
${\widetilde L}_N \equiv {\widetilde {\mathfrak L}} - \partial_{\tau}^2$,
acting on scalars in the spatial section of the manifold defined
by the metric Eq.~(\ref{newmetric}).
\begin{eqnarray}
{\widetilde {L}}_N & = & -\triangle_N^{{\widetilde g}}-\rho_N^2+
e^{2\tau}(m^2+\xi R^g)\:,
\label{da}
\\
\triangle_N^{{\widetilde g}} &=& \partial_{\tau}^2
- (N-1)\partial_{\tau} + e^{2\tau}\triangle_{N-1}\,,
\nonumber
\end{eqnarray}
where $\triangle_{N-1}$ is the Laplace-Beltrami operator on
$(N-1)-$dimensional space, $\rho_N=(N-1)/2$.
It should be noted the appearance of an effective ``tachionic'' mass
$-\rho_N^2$, which has important consequences on the structure of the
zeta function related to the operator ${\widetilde L}_N$, which
has generally speaking a continuum spectrum (see \cite{Bytsenko3}
for details).

\begin{proposition} (A. A. Bytsenko, G. Cognola and S. Zerbini
\cite{Bytsenko3})\,\,\,
The trace of the heat kernel has the form
\begin{eqnarray}
{\rm Tr} e^{-t{\widetilde L}_N} & = &
\sum_{n=0}^{(N-3)/2}
\frac{A_{2n}(L_{N-1})\,a(t|-\triangle_{{\mathbb H}^{N-2n}}
-\rho^2_{N-2n})}
{N-1-2n}(4\pi \varepsilon^{-2})^{(N-1-2n)/2}
\nonumber \\
& + & \frac{1}{4\sqrt{\pi t}}
\left[\frac{d}{ds}\zeta(s|L_{N-1})|_{s=0}
-2\zeta(0|L_{N-1}){\rm log}(\varepsilon/2)\right]
\nonumber \\
& - & \frac{1}{4}\zeta(0|L_{N-1})
+\frac{1}{2\pi}\zeta(0|L_{N-1})
\int_{\mathbb R}dr\psi(ir)e^{-tr^2}\:,
\label{Ktodd}
\end{eqnarray}
\begin{eqnarray}
{\rm Tr} e^{-t{\widetilde L}_N} & = & \sum_{n=0}^{(N-2)/2}
\frac{A_{2n}(L_{N-1})\,a(t|-\triangle_{{\mathbb H}^{N-2n}}
-\rho^2_{N-2n})}
{N-1-2n}(4\pi \varepsilon^{-2})^{(N-1-2n)/2}
\nonumber \\
& + & \frac{1}{4\sqrt{\pi t}}\frac{d}{ds}\zeta(s|L_{N-1})|_{s=0}
\:,
\label{Kteven}
\end{eqnarray}
valid for odd and even $N$ respectively.
Here by $a(t|-\triangle_{{\mathbb H}^{N-2n}}-\rho^2_{N-2n})$
we indicate the
diagonal heat kernel of a Laplace-like operator on ${\mathbb H}^{N-2n}$,
and $\varepsilon$ is a horizon cutoff parameter in integrating
over coordinates.
\end{proposition}

\section{Constant time slices in Milne cosmology}

Now let us consider eleven-dimensional metric
\begin{equation}
ds^2_{10} = -dt^2+t^2
d{\Bbb H}^N+dx_1^2+dx_2^2+dx_3^2+\sum_{j=1}^{7-N}dy_j^2 \ ,
\label{metric10}
\end{equation}
where the
$y$ coordinates describe compact internal dimensions.
This is an exact solution of M-theory \cite{Kehagias,Russo,Bytsenko00}.
The internal space described by the $y$ coordinates can be
replaced by any Ricci flat space, giving a more general class of
cosmological backgrounds. Note that four dimensional
Friedmann-Robertson-Walker cosmology can be
obtained from this model \cite{Russo}. First, we replace the hyperboloid
${\Bbb H}^N$ by a finite volume space $\Gamma\backslash {\Bbb H}^N$,
where
$\Gamma$ is a discrete subgroup of isometries such that
the space has finite
volume. Then we compactify to four dimensions.
To obtain the four dimensional
Einstein frame metric, we write the metric in the form
\begin{eqnarray}
ds^2 & = & e^{2a(t)}ds^2_{4E}+e^{2b(t)}d{\Bbb H}^N+
\sum_{j=1}^{7-N}dy_j^2\,,
\\
ds^2_{4E} & = & e^{2c(t)}(-dt^2+dx_1^2+dx_2^2+dx_3^2) \,\, .
\end{eqnarray}
If the condition $e^{2a}e^{Nb}=1$ is satisfied, then $ds^2_{4E}$
is the Einstein frame metric. Comparing to (\ref{metric10}), one
obtains
\begin{equation}
ds^2_{4E} =
t^N(-dt^2+dx_1^2+dx_2^2+dx_3^2)\ ,
\end{equation}
or
\begin{equation}
ds^2_{4E} =
-d \tau ^2+ \tau ^{2N/(N+2)}(dx_1^2+dx_2^2+dx_3^2).
\end{equation}
This corresponds to $4D$ Einstein equations coupled to an energy
momentum tensor of a perfect fluid with equation of state $
p=\kappa \rho ,\, \kappa=(4-D)/3D. $ Although we have started with
vacuum Einstein equations in ten dimensions, the four dimensional
Einstein metric describes a homogeneous and isotropic space in
presence of matter. This matter is, of course,  the scalar field
associated with the modulus representing the volume of the
hyperbolic space. Interestingly, the above metric is the
asymptotic (large time) form of the models of \cite{Townsend}. For
$D=4$, it describes a universe filled with dust ($p=0$), and for
$D>4$ a universe filled with negative pressure matter\footnote{In
passing, we mention that, in Einstein gravity, the accelerated
expansion of a spatially flat universe, requires the cosmological
dynamics to be dominated by some exotic matter with negative
pressure. We plan to address to this problem in a forthcoming
paper.}.

Since the models are based on a flat eleven-dimensional geometry, the
$(N+1)-$ dimensional Milne universes provide a simple setup for
the study of interesting cosmological models. Let us
consider strings/branes propagating in this space. An important question
is whether the model is exactly solvable. To start with,
consider the model based on ${\Bbb H}^N$ with no identification,
i.e. $\Gamma$ is trivial. From the relation
$t^2 - UV+\sum_{j=1}^{N-1}X_j^2 = 0$ it follows that
$ UV-X_j^2\geq 0.$ If the physical space is restricted
only to this Milne patch,
say with $t>0$, then the brane coordinates are subject to the
constraint that the brane lives in the interior of the future
directed light cone; the space is not geodesically complete and a
full description requires boundary conditions at the light cone
surface. If it is possible that consistency also requires the
inclusion of the past light cone (in string theory it is
possible), then the geometry would describe a universe contracting
to a big crunch which makes a transition to an expanding big bang
universe. It is non-trivial to impose the condition as
$UV-X_j^2\geq 0$ in brane theory. On the other hand, if the full
space $U,V,X_j$ is considered, closed timelike curves can arise in
the exterior of light cone as a result of identifications.

\section{Hyperbolic geometry in M--theory}

Let us consider an irreducible rank one symmetric space
$X = G/K$ of non-compact type. Thus $G$ will be a connected
non-compact simple split rank one Lie group with finite center and
$K\subset G$ will be a maximal compact subgroup.
Up to local isomorphism we can represent $X$ by the following
quotients:
\begin{eqnarray}
X = && SO_1(N,1)/SO(N)\,,\,\,\, SU(N,1)/U(N)\,,\,\,\,
\nonumber \\
&& SP(N,1)/(SP(N)\times SP(1))\,,\,\,\, F_{4(-20)}/Spin(9)\,,
\end{eqnarray}
where the dimension of the spaces is $N, 2N, 4N, 16$
respectively in these cases. For details on these matters
the reader may consult \cite{Helgason}.
The spherical harmonic analysis on $X$ is controlled by
Harish-Chandra's Plancherel density $\mu (r)$, a function on
the real numbers $\Bbb R$, computed by Miatello
\cite{Miatello1,Miatello3,Bytsenko11}, and others, in
the rank one case we are considering.
The object of interest is the groups $G=SO_1(N,1)$ \,$(N\in {\Bbb
Z}_{+})$ and  $K=SO(N)$. The corresponding symmetric space of
non-compact type is the real hyperbolic space $X = {\Bbb H}^N =
SO_1(N, 1)/SO(N)$ of sectional curvature $-1$. Its compact dual
space is the unit $N-$sphere.

\subsection{Co-compact group}

Let $\tau$ be an irreducible representation of $K$ on a complex
vector space $V_\tau$, and form the induced homogeneous vector
bundle $G\times_K V_\tau$. Restricting the $G$ action
to $\Gamma$ we obtain the quotient bundle $E_\tau=\Gamma\backslash
(G\times_KV_\tau)\longrightarrow X_{\Gamma}=\Gamma\backslash X$
over $X$.
The natural Riemannian structure on $X$ (therefore on $X_{\Gamma}$)
induced
by the Killing form $(\;,\;)$ of $G$ gives rise to a connection
Laplacian ${L}_{\Gamma}$ on $E_\tau$. If $\Omega_K$ denotes the
Casimir operator of $K-$that is $ \Omega_K=-\sum y_j^2, $ for a
basis $\{y_j\}$ of the Lie algebra ${\mathfrak k}_0$ of $K$, where
$(y_j\;,y_\ell)=-\delta_{j\ell}$, then
$\tau(\Omega_K)=\lambda_\tau{\mathbf 1}$ for a suitable scalar
$\lambda_\tau$. Moreover for the Casimir operator $\Omega$ of $G$,
with $\Omega$ operating on smooth sections $\Gamma^\infty E_\tau$
of $E_\tau$ one has
$
{L}_{\Gamma}=\Omega-\lambda_\tau{\mathbf 1}\,.
$
For $\lambda\geq 0$ let
$
\Gamma^\infty\left(X_{\Gamma}\;,E_\tau\right)_\lambda=
\left\{s\in\Gamma^\infty E_\tau\left|-{L}_{\Gamma}s=\lambda s\right.
\right\}
$
be the space of eigensections of ${L}_{\Gamma}$ corresponding to
$\lambda$. Here we note that if $X_{\Gamma}$ is compact we can
order the spectrum of $-{L}_{\Gamma}$ by taking
$0=\lambda_0<\lambda_1<\lambda_2<\cdots$;
$\lim_{j\rightarrow\infty}\lambda_j=\infty$.

\begin{theorem} (A. A. Bytsenko and F. L. Williams \cite{byts2})
\,\,\, The heat kernel admits an asymptotic
expansion (\ref{kernel}), and for all $G$ except $G=SO_1(\ell,1)$ with
$\ell$ odd, and for $0\leq k\leq N/2-1$,
\begin{equation}
A_k(L_{\Gamma}) = (4\pi)^{\frac{N}{2}-1}\chi(1){\rm Vol}
(\Gamma\backslash G)C_G\pi
\sum_{\ell=0}^k
\frac{(-\rho_N^2)^{k-\ell}}{(k-\ell)!}\Bigr[\frac{N}{2}-(\ell+1)\Bigr]!
a_{2[\frac{N}{2}-(\ell+1)]}
\mbox{,}
\end{equation}
while for $n=0,1,2,...$ we have
\begin{eqnarray}
A_{\frac{N}{2}+n}(L_{\Gamma}) & = & (-1)^n(4\pi)^{\frac{N}{2}-1}\chi(1)
{\rm Vol}(\Gamma\backslash G)C_G\pi
\left[\sum_{j=0}^{\frac{N}{2}-1}(-1)^{j+1}\frac{\rho_N^{2(n+1+j)}j!a_{2j}}
{(n+1+j)!} \right.
\nonumber \\
& + & \left. 2\sum_{j=0}^{\frac{N}{2}-1}\sum_{\ell=0}^n(-1)^{\ell}
\frac{\rho_0^{2(n-\ell)}}{(n-\ell)!}\beta_{\ell+1}(j)a_{2j}\right]
\mbox{.}
\end{eqnarray}
Here $\beta_r(j)\,\,(r\in {\Bbb Z}_{+})$ is given by
\begin{equation}
\beta_r(j)\stackrel{def}{=}\left[2^{1-2(r+j)}-1
\right]\left[\frac{\pi}{a(G)}
\right]^{2(r+j)}
\frac{(-1)^jB_{2(r+j)}}{2(r+j)[(r-1)!]}
\mbox{,}
\end{equation}
$B_r$ is the $r$-th Bernoulli number,
$
a(G)\stackrel{def}{=}\pi
$
if $G=SO_1(\ell,1)$ with $\ell$ even, and $a_{2j},\, C_G$
are some constants ($C_G$ depending on $G$).
For $G=SO_1(2n+1,1),\,\, k=0,1,2, ...$
\begin{equation}
A_k(L_{\Gamma})=\pi(4\pi)^{n-\frac{1}{2}}\chi(1){\rm Vol}
(\Gamma\backslash G)C_G
\sum_{\ell=0}^{{\rm min}(k,n)}
\frac{(-n^2)^{k-\ell}\Gamma\left(n-\ell+\frac{1}{2}\right)a_{2(n-\ell)}}
{(k-\ell)!}
\mbox{.}
\end{equation}
\end{theorem}

\subsection{The orbifold coset:
$\Gamma = SL(2,{\Bbb Z}+i{\Bbb Z})/\{\pm Id\}$}

In \cite{Russo} the $SL(2,{\mathbb Z})$ orbifold model from Milne
spaces and the string spectrum associated with that orbifold has
been analysed. It has been also shown that strings with
$SL(2,{\mathbb Z})$ identifications are related to the null
orbifold \cite{Horowitz1} with an extra reflection generator.

Here we consider the case $N=3$ and the group of
local isometry associated with a simple three-dimensional complex
Lie group. The discrete group can be chosen in the form $\Gamma
\subset PSL(2, {\mathbb C})\equiv SL(2,\mathbb C)/\{\pm Id\}$,
where $Id$ is the $2\times 2$ identity matrix and is an isolated
element of $\Gamma$. The group $\Gamma$ acts discontinuously at
point $z\in\bar{\mathbb C}$, $\bar{\mathbb C}$ being the extended
complex plane. We consider a special discrete group $SL(2,{\mathbb
Z}+i{\mathbb Z})/\{\pm Id\}$, where $\mathbb Z$ is the ring of
integer numbers. The element $\gamma\in\Gamma$ will be identified
with $-\gamma$. The group $\Gamma$ has, within a conjugation, one
maximal parabolic subgroup $\Gamma_\infty$.
Let us consider an arbitrary integral operator with kernel
$k(z,z')$. Invariance of the operator is equivalent to fulfillment
of the condition
$k(\gamma z,\gamma z') = k(z,z')$ for any
$z,z'\in {\mathbb H}^3$ and $\gamma\in
PSL(2,{\mathbb C})$. So the kernel of the invariant operator is a
function of the geodesic distance between $z$ and $z'$. It is
convenient to replace such a distance with the fundamental
invariant of a pair of points $u(z,z')=|z-z'|^2/yy'$, thus
$k(z,z')=k(u(z,z'))$ . Let $\lambda_j$ be the isolated eigenvalues
of the self-adjoint extension of the Laplace operator and let us
introduce a suitable analytic function $h(r)$ and
$r^2_j=\lambda_j-1$. It can be shown that $ h(r)$ is related to
the quantity $k(u( z,\gamma z))$ by means of the Selberg
transform. Let us denote by $g(u)$ the Fourier transform of $
h(r)$, namely $g(u)= (2\pi)^{-1}\int_{\mathbb
R}dr\,h(r)\exp(-iru)$.

\begin{theorem}
Suppose $h(r)$
be an even analytic function in the strip $|\Im r|<1+\varepsilon $
($\varepsilon>0$), and $h(r)={\mathcal O}(1+|r|^2)^{-2}$.
For the special discrete group
$SL(2,{\mathbb Z}+i{\mathbb Z})/\{\pm Id\}$
the Selberg trace formula holds
\begin{eqnarray}
\sum_j h(r_j) & - &
\sum_{\scriptstyle\{\gamma\}_{\Gamma},\gamma\not=Id,
\atop\scriptstyle\gamma-non-parabolic}\int d\mu(z)\, k(u(z,\gamma
z))
\nonumber \\
& - & \frac{1}{4\pi}\int_{\mathbb R} dr\, h(r)
\frac{d}{ds}{\log}\,S(s)|_{\atop s=1+ir} + \frac{h(0)}{4}[S(1) -1]
-{\mathfrak C}g(0)
\nonumber \\
& = & {\rm Vol}(\Gamma\backslash G)\!\int_0^\infty \! \frac{dr\,
r^2}{2\pi^2}\:h(r) -\frac{1}{4\pi}\!\int_{\mathbb R}\!\! dr h(r)
\psi(1+ir/2)
\mbox{.}
\label{Selberg}
\end{eqnarray}
\end{theorem}

The first term in the right hand site of
Eq.~(\ref{Selberg}) is the contribution of the identity element,
${\rm Vol}(\Gamma\backslash G)$ is the (finite) volume of
the fundamental domain with respect to the measure $d\mu$,
$\psi(s)$ is the logarithmic derivative of the Euler $\Gamma-$function,
and ${\mathfrak C}$ is a computable real constant
\cite{Bytsenko1,Bytsenko,Bytsenko2}.
The function $S(s)$ is given by a generalised Dirichlet series
$S(s)=\pi^{1/2}\Gamma(s-1/2)[\Gamma(s)]^{-1} \sum_{c\neq
0}\sum_{0\leq d<|c|} |c|^{-2s}$, where the sums are taken
over all pairs $c,d$ of the matrix
$\left(\begin{array}{ll}*\,\,*\\
c\,\,d\end{array}\right)\subset\Gamma_\infty
\backslash\Gamma/\Gamma_\infty$.
The meromorphic function $S(s)$ convergent for $\Re\,s>1$, and
it poles are contained in the region $\mathop{\Re}\nolimits
s<1/2$ and in the interval $[1/2,1]$.
\\
\\
In general, the determinant of an elliptic differential operator
requires a regularization. It is convenient to introduce the
operator ${L}_{\Gamma}(\delta)= {L}_{\Gamma} + \delta^2-1$,
with $\delta$ a suitable parameter. One of the most
widely used regularization is the zeta-function regularization.
Thus, one has ${\rm log} {\rm det}\, {L}_{\Gamma} (\delta)=
-(d/ds)\zeta ( s|{L}_{\Gamma} (\delta))|_{(s=0)}$.
In standard cases, the zeta
function at $s=0$ is well defined and  one gets a finite result.
The meromorphic structure of the analytically continued zeta
function is related to the asymptotic properties of the
heat-kernel trace. Summarizing, the final result is:

\begin{theorem} (A. A. Bytsenko, G. Cognola and S. Zerbini
\cite{Bytsenko}) \,\,
The following identity holds
\begin{equation}
\det {L}_{\Gamma} (\delta) =
\frac{2}{(\pi\delta)^{1/2}\Gamma(\delta/2)}
\exp\left(-\frac{1}{6\pi} {\rm Vol}(\Gamma\backslash G)
\delta^3+{\frak C}\delta\right)\,
Z_{\Gamma}(1+\delta)
\mbox{,}
\end{equation}
where $Z_{\Gamma}(s)$ is the Selberg's zeta function.
\end{theorem}

Let us analyse a scalar field propagating in this orbifolds.
Normalizable wave functions associated with scalar density
can be written in terms of cusp forms. Cusp forms are authomorphic
functions which decrease exponentially at infinity. The discrete
part of spectrum associate with cusp forms, while the Eisenstein
series related to the continuous part. A vertex operator of
a brane model contains cusp forms. In the string case a
computation of $S-$matrix elements by using plane-wave vertex
operators has been discussed in \cite{Russo}. Such computation
for Kaluza-Klein quantum numbers of brane modes turn out to be
more complicate and we disregard it.

Finally we note that in the $SL(2,{\Bbb Z}+ i{\Bbb Z})/\{\pm Id\}$
orbifold
the instability may be absent. To demonstrate that for three-orbifold
we can use the arguments given for string models in \cite{Russo}.
The instability can be originated from the gravitational interaction
of plane waves.
The continuum part of spectrum may lead to wave interactions, but it
is severely restricted by $SL(2,{\Bbb Z} + i{\Bbb Z})/\{\pm Id\}$
symmetry,
and the argument of \cite{Horowitz} on instability
does not seem to directly apply to our case.
In the discrete part of the spectrum the states have finite motion,
and the corresponding wave functions are regular.

\section{Cosets $\Gamma\backslash G/K$ and Killing spinors}

In the previous sections we have considered real hyperbolic space
forms. The hyperbolic spaces ${\Bbb H}^N$ have Killing spinors
transforming in the spinorial representation of $SO_1(N-1,1)$
\cite{Fujii} (see also \cite{Kehagias,Lu,Nasri}). Thus the
simplest membrane model with trivial $\Gamma$ allows
supersymmetry. In general, the following results hold:

\begin{proposition} (T. Friedrich \cite{Friedrich})\,\,
A Riemannian spin
manifold $(M^N, g)$ admitting a Killing spinor $\psi \neq 0$ with
Killing number $\mu \neq 0$ is locally irreducible.
\end{proposition}
\noindent
{\bf Proof}.
Let
the locally Riemannian product has the form $M^N = M^K\times
M^{N-K}$. Let ${\mathcal X}, {\mathcal Y}$ are vectors tangent to $M^K$
and $M^{N-K}$ respectively, and, therefore, the curvature tensor
of the Riemannian manifold $(M^N, g)$ is trivial. Since $\psi$ is
a Killing spinor the following equations hold:
\begin{eqnarray}
\nabla_{\mathcal X}\psi & = & \mu {\mathcal X} \cdot \psi,
\nonumber \\
4\mu^2 & = & [N(N-1)]^{-1}R\,\,\,\, {\rm at}\,\,\,{\rm each}
\,\,\,{\rm point}\,\,\,{\rm of}\,\,\, {\rm a}\,\,\,
{\rm connected}\,\,\, {\rm Riemannian}
\nonumber \\
&& {\rm spin}
\,\,\, {\rm manifold} \,\,\, (M^N, g)
\label{Kil}
\mbox{,}
\end{eqnarray}
where $R$ is a scalar curvature. Because of (\ref{Kil})
we have
\begin{eqnarray}
&& \nabla_{{\mathcal X}} \nabla_{\mathcal Y}\psi =
\mu(\nabla_{\mathcal X} {\mathcal Y})
\cdot\psi+\mu^2{\mathcal Y}\cdot{\mathcal X}\cdot \psi
\Longrightarrow
\nonumber \\
&& (\nabla_{\mathcal X}\nabla_{\mathcal Y} -
\nabla_{\mathcal Y}\nabla_{\mathcal X}
- \nabla_{[{\mathcal X}, {\mathcal Y}]})\psi =
\mu^2({\mathcal Y}\cdot{\mathcal X} -{\mathcal X}\cdot{\mathcal Y})\psi
\label{nabla}
\mbox{.}
\end{eqnarray}
The curvature tensor $R({\mathcal X},{\mathcal Y})$ in the spinor
bundle ${\mathfrak S}$ is related to the curvature tensor of the
Riemannian manifold $(M^N, g)$:
$
R({\mathcal X},{\mathcal Y}) = (1/4)\sum_{j=1}^{N}e_j
R({\mathcal X},{\mathcal Y})e_j\cdot \psi,
$
where $\{e_j\}_{j=1}^N$ is a orthogonal basis in manifold.
Therefore Eq. (\ref{nabla}) can also be written as
\begin{equation}
\sum_{j=1}^{N}e_j R({\mathcal X},{\mathcal Y})e_j\psi
+[N(N-1)]^{-1}R
({\mathcal X}{\mathcal Y} - {\mathcal Y}{\mathcal X})\psi = 0
\label{curv}
\mbox{.}
\end{equation}
From Eq. (\ref{curv}) we get $R\cdot {\mathcal X}\cdot {\mathcal
Y}\cdot \psi =0$, and moreover ${\mathcal X}$ and ${\mathcal Y}$
are orthogonal vectors. Since $\mu \neq 0$ ($R \neq 0$) it follows
that $\psi =0$, hence a contradiction. ${\Box}$
\\
We have also the following statement:
\begin{proposition} (T. Friedrich \cite{Friedrich})
Let $(M^N,g)$ be a connected Riemannian spin manifold and let
$\psi$ is a non-trivial Killing spinor with Killing number
$\mu \neq 0$. Then $(M^N,g)$ is an Einstein space.
\end{proposition}
{\bf Proof}. The proof easily follows from Proposition 2; indeed
$(M^N,g)$ is an Einstein space of scalar curvature given by Eq.
(\ref{Kil}). $\Box$

There are no normalizable
modes for any field configurations in hyperbolic spaces.
Spaces with finite volume of fundamental domain can be obtained
by forming the coset spaces with topology
$\Gamma\backslash {\mathbb H}^N$ where $\Gamma$ is a discrete
subgroup of the isometry group.
Let us comment on the supersymmetry of these spaces following the
lines of \cite{Kehagias,Russo}.
For non-trivial $\Gamma$ and finite volume space
$\Gamma\backslash {\Bbb H}^N$ it has been shown \cite{Kehagias}
that for even $N$ supersymmetries are always broken by the
identifications.
Indeed, the isometry group of ${\mathbb H}^N$ is $SO_1(N,1)$ and
$\Gamma$ is in general a subgroup of $SO_1(N,1)$, which may or
may not have fixed points. Killing spinors are in the spinorial
representation of $SO_1(N-1,1)$, and if $\Gamma$ is a subgroup of
$SO_1(N-1,1)$, but it is not a subgroup of $SO_1(N-3,1)$, then there
are no surviving Killing spinors. The later exist if
$\Gamma \in SO_1(N-3,1)$, but in this case
$\Gamma\backslash{\mathbb H}^N$ will still be of infinite volume.
Therefore, for even $N$ there is no finite volume cosets
$\Gamma\backslash{\mathbb H}^N$ with unbroken supersymmetries.
On the other hands, for odd $N$ this analysis
does not exclude that an appropriate choice of $\Gamma$
could give a supersymmetric model with finite volume hyperbolic
space.
For odd $N$ there are two Killing spinors on ${\mathbb H}^N$
in the spinorial representation of $SO_1(N-1,1)$. These
spinors are also Weyl spinors of the isometry group $SO_1(N,1)$,
so they form an irreducible Dirac spinor of $SO_1(N,1)$.
All supersymmetries are broken if $\Gamma$ is not a
subgroup $SO_1(N-1,1)$. If $\Gamma$ is a subgroup $SO_1(N-1,1)$,
then half of the supersymmetries survive.
A question of interest is whether supersymmetry
survives under the orbifolding by the discrete group $\Gamma$.
Perhaps there are more solutions involving real hyperbolic
spaces, where some supersymmetries are unbroken.
However the analysis of that problem is complicated and we leave it
for other occasion.

\subsection*{Acknowledgements}

A. A. Bytsenko and M. E. X. Guimar\~aes would like to thank the
Conselho Nacional de Desenvolvimento Cient\'{\i}fico e
Tecnol\'ogico (CNPq) for a support. The authors would like to
thank the Coordenacao de Campos e Part\'{\i}culas do Centro
Brasileiro de Pesquisas F\'{\i}sicas (CCP/CBPF) for kind
hospitality during preparation of this work.

\end{document}